\documentclass{emulateapj}

\usepackage{natbib}
\usepackage{bm}

\bibstyle{apj}

\newcommand\arcsec{\mbox{$^{\prime\prime}$}}

\begin{document}

\author{Neal Dalal\altaffilmark{1} and Gilbert Holder} 
\affil{Institute for Advanced Study, Einstein Drive, Princeton NJ 08540}
\author{Joseph F.\ Hennawi} 
\affil{Department of Astrophysical Sciences, Princeton University, 
Ivy Lane, Princeton NJ 08544}

\altaffiltext{1}{Hubble Fellow}

\title{Statistics of Giant Arcs in Galaxy Clusters}

\begin{abstract}
  We study the expected properties and statistics of giant arcs
  produced by galaxy clusters in a $\Lambda$CDM universe and 
  investigate how the characteristics of CDM clusters determine the
  properties of the arcs they generate.  Due to the triaxiality and
  substructure of CDM halos, the giant arc cross section for
  individual clusters varies by more than an order of magnitude as a
  function of viewing angle.  In addition, the shallow density cusps
  and triaxiality of CDM clusters cause systematic alignments of
  giant arcs which should be testable with larger samples from
  forthcoming lensing surveys.  We compute the predicted
  statistics of giant arcs for the $\Lambda$CDM model and compare to
  results from previous surveys.  The predicted arc
  statistics are in excellent agreement with the numbers of giant arcs
  observed around low redshift ($0.2\lesssim z\lesssim 0.6$) clusters
  from the EMSS sample,
  however there are hints of a possible excess of arcs observed around
  high redshift $z\gtrsim 0.6$ clusters.  This excess, if real,
  appears to be due to the presence of highly massive or concentrated
  clusters at high redshifts.
\end{abstract}


\maketitle

\section{Introduction}

Clusters of galaxies provide some of the most spectacular examples of
gravitational lensing. The popular images of Abell 2218 and
CL0024+1654 that are prominent on the Hubble Space Telescope public
website (www.hubblesite.org) show very elongated multiply imaged
background galaxies whose features have been distorted by the deep
gravitational potential of the galaxy cluster.  Such cluster lenses
have a variety of cosmological uses.  For example, these natural
gravitational telescopes can greatly magnify distant sources, allowing
us to study the properties of otherwise unobservable galaxies
\citep[e.g.][]{blain99,smail02,metcalfe03}.  
In addition, strongly lensed arcs
offer a unique, direct probe of the cluster gravitational potential on
scales where dark matter is expected to be the dominant
component. Heating, cooling, and star formation can have a significant
impact on baryons but the dark matter, and therefore the total
potential, should be relatively unaffected by these poorly understood
processes.  Lastly, the incidence of giant arcs may be used to study
the background cosmology itself.  \citet[hereafter B98]{bartelmann98}
found that the predicted number of giant arcs varies by orders of
magnitude among different cosmological models.  In light of mounting
evidence supporting the $\Lambda$CDM model (e.g. Spergel et al. 2003),
\nocite{spergel03}
the spectacular images of giant arcs have led to an embarrassment of
riches, as B98 found that the observed instances of giant arcs
exceeded their predicted rate by an order of magnitude.

The original discrepancy reported in B98 was based on a subsample of
16 of the most massive clusters \citep{lefevre94} in the EMSS survey and
the arc frequency (roughly 20\% of very massive clusters) 
was confirmed in the larger EMSS sample of
\citet{luppino99}. A subsequent survey based on 
optically-selected clusters in the Las Campanas Distant Cluster
Survey \citep{zaritsky03} found comparable giant arc frequencies, while
a recent report by the Red Cluster Sequence cluster survey group
\citep{gladders02,gladders03} found a high
probability ($\sim 30\%$) of a lensing cluster showing
more than one source being distorted into a giant arc.

On the theoretical side, significant work has gone into refining the
expected number of giant arcs. Several works have confirmed the
lensing cross-sections of B98 
\citep{meneghetti01,meneghetti03a,bartelmann02}, but other recent
work has suggested that the cross-sections of B98 may be too low.
In particular \citet{wambsganss03} used numerical
simulations and found that the cross-section was a very steep function
of redshift and that the anomalous cross-sections could be brought into
agreement by allowing a broader range of source redshifts. \citet{oguri03} 
used analytic models with triaxiality and found
that allowing a steeper central density profile enhanced the lensing
cross-section to a level that was close to the observed lensing frequency
of the EMSS sample. 

In this paper we go back to the beginning of the problem (i.e., B98)
to repeat the analysis on a larger sample of simulated clusters and
update the statistics for the EMSS lensing sample. We then apply the
analysis to recent optical catalogs to determine the extent, if any,
of the discrepancy between observations and theory.  Lastly, we
investigate what new information may be gleaned from upcoming deep,
large area surveys like the CFHT Legacy survey, which should produce
large numbers of new strongly lensed arcs.

\section{Cluster Lensing Simulations}

We estimate giant arc statistics by ray-tracing through cosmological
N-body simulations.  Below, details of the calculation are briefly
described.

\subsection{N-body simulations}

We used the publicly available simulation outputs of the Virgo 
Project\footnote{http://www.mpa-garching.mpg.de/Virgo}. We used the
$\Lambda$ CDM simulations that were performed as part of the GIF project
\citep{kauffmann99}. 
The cosmological parameters were $\Omega_m=0.3$, $\Omega_{\Lambda}=0.7$,
$h=0.7$, power spectrum shape parameter $\Gamma=0.21$, and mass power
spectrum normalization $\sigma_8=0.9$. The comoving simulation box length
was 141.3 $h^{-1}$ Mpc and $256^3$ particles were used, leading to a mass
per particle of $10^{10} h^{-1} M_\odot$. The gravitational softening
length was $20 h^{-1}$kpc. 
Simulation outputs were available at a variety of redshifts, allowing
accurate allowance of time evolution in cluster properties. 
A subset of the clusters from these simulations were used as part of the
ensemble of simulated clusters used in B98.

\subsection{Lensing calculation}
To compute giant arc statistics for the N-body simulations described
above, we adopt a technique similar to that pioneered by
\citet{bartelmann94} and employed by subsequent workers
\citep[e.g][]{bartelmann98,meneghetti00,meneghetti01,meneghetti03cd}.
We first project the dark matter density onto a 
two dimensional regular grid.  Typically we used a $256\times256$ grid
covering a square $4 h^{-1}$ comoving Mpc on a side, centered on the
cluster.  From the projected surface density $\Sigma$, the convergence
for each lens and source redshift pair is given by $\kappa =
\Sigma/\Sigma_{\rm crit}(z_l,z_s)$, where $\Sigma_{\rm crit}$ is the
critical surface density for strong lensing \citep{sef}.
Then, the deflection angle map ${\bm\alpha}({\bm x})$ is constructed
(via Fourier transform) on the same grid.   Given this map, we then
ray trace a finer $4000\times4000$ grid covering a subset of the image
plane onto the source plane, linearly interpolating the deflection
between the grid points of the coarser $256^2$ grid.  We set up a regular
grid on the source plane; each source plane pixel contains a linked
list of all image plane pixels which ray trace to the source plane
pixel.  Having performed this single ray trace, it is then a simple
matter to construct the lensed arcs for an arbitrary source by
concatenating the lists for all source plane pixels falling in the
source.  We emphasize that the use of linked lists greatly speeds the
calculation -- only one single ray trace is performed, after which
arcs may be constructed for many different source positions, sizes and
geometries for essentially no computational cost.

\begin{figure}
\plottwo{src.eps}{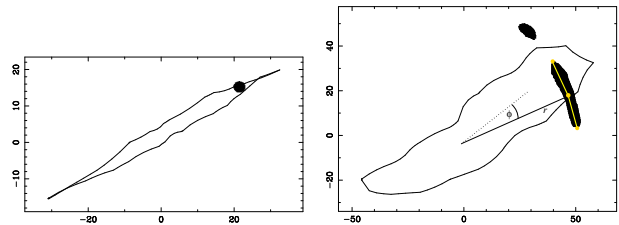}
\caption{Example of simulated arcs.  The source plane (left) and image
  plane (right) are illustrated.  Solid lines depict the approximate
  locations of the critical lines and caustics, while filled regions
  show the pixels falling inside the source (left panel) and the
  corresponding lensed images (right panel).  Two arcs are found for
  this source, with length to width ratios of 7.5 and 2.8
  respectively.  For the long arc, the arc radius $r$ and position
  angle $\phi$ (relative to the major axis of the mass distribution)
  are labeled.
  \label{fig:arcexample}
}
\end{figure}

For each source, the pixel lists are then sorted into separate arcs,
by grouping together neighboring pixels.  We measure arc properties
using methods based on \citet{bartelmann94}.  The arc area and
magnification are found by summing the areas of the pixels falling in
the arc.  Arc lengths are estimated by first finding the arc centers,
then finding the arc pixel furthest from the centroid, as well as the
pixel furthest from this pixel.  As in B98, the arc length was then
given by the sum of the lengths of the two line segments connecting
these three points.  The arc width was defined as the ratio of the arc
area to the arc length.  See figure \ref{fig:arcexample} for an
example.  

\begin{figure}
\plotone{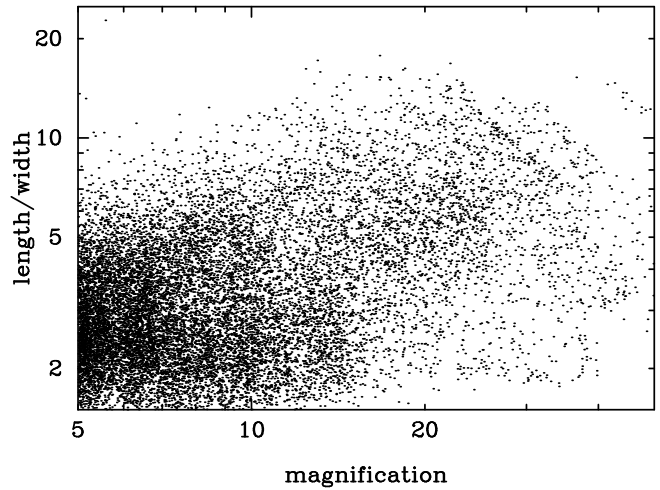}
\caption{Magnification vs.\ length/width ratio.  While these two
  quantities are equal for isothermal lenses, giant arcs tend to form
  at small radii ($r<r_s$) where the density profile is shallower than
  isothermal.  Here we plot all arcs in one cluster projection with
  magnification $\mu>5$, however only $\sim 40\%$ of the points have
  length/width ratios $r>5$.  
\label{wbo}
}
\end{figure}

The relation between length/width ratio and magnification is sensitive
to the slope of the halo potential \citep{williams98} and can also
be affected by local perturbations in the projected mass distribution.
In figure~\ref{wbo} we plot arc magnification against
length/width ratio using the above method. We find good agreement
for the mean trend of length/width vs. magnification
with the predictions of \citet{williams98} for typical clusters
in the simulations. However, there is substantial scatter in this relation
that is likely caused by random source orientations and local
fluctuations in the surface mass density.  
It is apparent that highly distorted images will also be highly
magnified, but the converse is not true.

For each cluster projection, we compute giant arc statistics using a Monte
Carlo approach.  We randomly place elliptical sources, with axis ratios drawn
uniformly in the range $q\in[0.5,1]$ and random position angles.  We
count as giant arcs only those images with length/width ratios
$r\geq10$, making no distinction between radial and tangential arcs,
although we note that the latter dominate the cross section.  Cross
sections are then computed by multiplying the fraction of sources
producing giant arcs in the Monte Carlo by the area over which the
sources were randomly distributed.  

\subsection{Cross sections}
\begin{figure}
\plotone{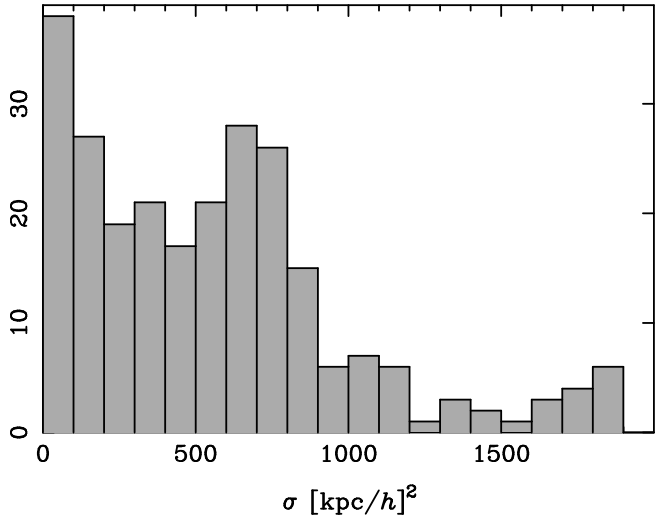}
\caption{Histogram of cross-sections for the 4$^{\rm th}$ most massive
  cluster at $z=0.2$ under different projections.
\label{fig:sighist}
}
\end{figure}

The giant arc cross section for a single cluster varies
enormously as a function of orientation, as shown in fig.~\ref{fig:sighist}.  
A factor of $\sim 20$ spread in $\sigma$ is not uncommon.
Additionally, the distribution is quite skewed, with a tail extending
to high $\sigma$.  
This appears to be due to the large degree of substructure 
and ellipticity/triaxiality in the density profiles.  Orientations
with large cross sections are typically those which project along the
cluster major axis, or which project massive substructures onto small
radii.  Unfortunately, this means that shot noise will be persistent
in any calculation of arc statistics; large numbers of projections are
required for trustworthy mean cross sections.  In the calculations
below, we compute mean cross sections by averaging over 251
orientations for each of the top 20 most massive clusters at each
redshift, while we perform 3 projections each for the next 100
clusters.

\begin{figure*}
\centerline{\includegraphics[width=0.8\textwidth]{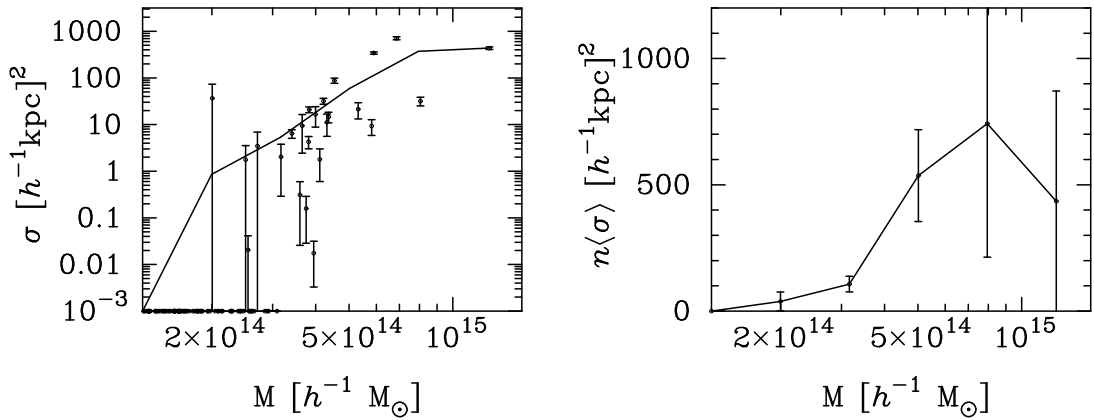}}
\caption{
Cross-section for forming giant arcs (left) and cross-section
weighted by number of clusters (right) for simulation volume at 
$z_l=0.2$ and a source redshift $z_s=1$.  Cross sections are given in
physical distances on the source plane. The solid line in the left
panel is the mean cross section $\langle\sigma\rangle$ averaged over
bins of width 0.2 in $\log_{10}M$. Error bars in the left panel
show error in the mean cross-section per cluster, not the dispersion,
while the error bars in the right panel include the effect of Poisson
statistics. 
\label{fig:cross-section}
}
\end{figure*}

In addition to the scatter in cross section for individual
clusters, there is large scatter among cross sections of
clusters of similar mass, shown in Figure~\ref{fig:cross-section}.
The error bar for each cluster shows the error in the mean, averaged
over all orientations. The dispersion for each cluster is comparable to
the mean.  The mean lensing cross-section as a function of mass shows a dramatic
rise at a mass of a few $10^{14} \, h^{-1} M_\odot$ and an apparent
flattening at the highest masses; the numbers become too low to 
say anything definite at high masses, as shown by the large error bar
in the highest mass bin.  

When weighted by the number of clusters at each
mass (right panel) there is an apparent peak at 
$\sim 7 \times 10^{14} h^{-1} M_{\odot}$. The significance of this
peak is hard to judge; given the low numbers of clusters in the
highest mass bin, and the large scatter among cross sections in each
bin, we cannot definitively say whether we are underestimating the
total optical depth by neglecting massive clusters too rare to appear
in our limited simulation volume.  On the other hand, the peak 
corresponds exactly to
the mass range expected of massive X-ray luminous galaxy clusters
which appear to dominate the giant arc optical depth observed by
surveys like EMSS.  In addition, we know that the rise of
$\langle\sigma\rangle$ with $m$ must flatten at high masses, once the
Einstein radii approach the NFW scale radius $r_s$.  This ensures
that, at some high mass, the exponentially declining mass function
$n(m)$ will overpower the growing $\langle\sigma\rangle(m)$.  
We will assume that our simulation volume provides
a fair sample of the objects causing most of the lensing, with the
caveat that our estimates might plausibly be low by a factor of $\sim 2$.

A related caveat is that, since the total optical depth
is dominated by the most massive clusters, 
great caution must be employed before deriving cosmological
constraints from giant arc statistics.  While the abundance of such
clusters depends in part on cosmological parameters like
$\Omega_\Lambda$, it also depends sensitively upon the
matter power spectrum \citep[e.g.][]{bahcall98,haiman01}.  Strong prior
constraints on parameters like $\sigma_8$ must be applied (and
believed) before constraints can be derived on parameters like
$\Omega_\Lambda$ or $w$.

\subsection{Arc properties and cluster galaxies}
\label{galaxy}

The effects of cluster galaxies on the formation of giant arcs have
been investigated by \citet{meneghetti00} and \citet{flores00}, 
and the effects of central cD galaxies in particular were studied by
\citet{meneghetti03cd}.  These authors found that cluster galaxies 
are generally unimportant, except for central cD galaxies which
can enhance the giant arc cross section by $\sim 50\%$. 

\begin{figure}
\plotone{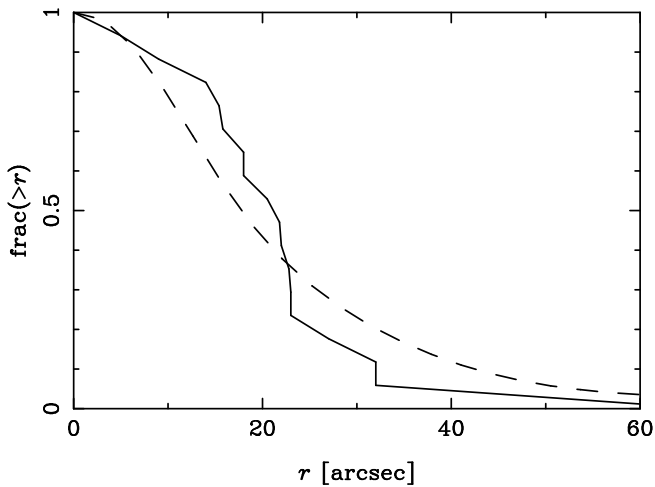}
\caption{Radial distribution of giant arcs, as measured by
  \citet{luppino99} (solid line), and as predicted by the $z=0.3$ GIF
  clusters (dashed line).
  \label{fig:raddist}
}
\end{figure}

\begin{figure}
\plotone{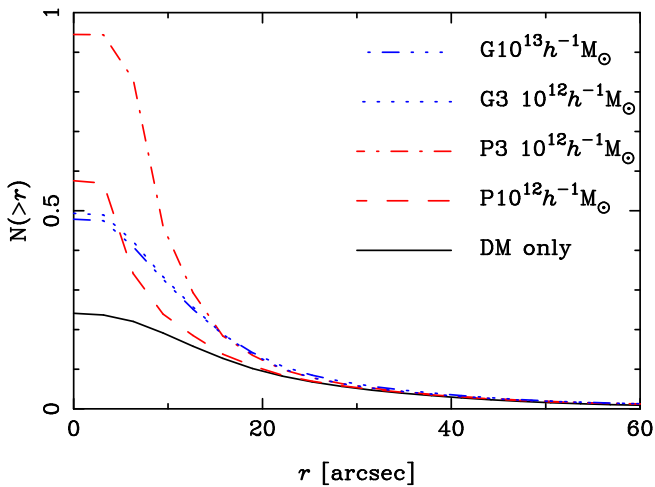}
\caption{Number of $z_s=2$ giant arcs produced by a (141 $h^{-1}$Mpc)$^3$
  volume of clusters at $z=0.3$.  The solid line shows results from
  the dark matter distribution alone, the `P' curves show results
  when a central point mass is added, and the `G' curves show the
  effects of adding a central singular isothermal sphere with velocity
  dispersion $\sigma=350$ km/s.
  \label{fig:withgal}
}
\end{figure}

The relatively small effect of cluster galaxies on the arc formation
cross section is not surprising.  Giant arcs typically form at large
radii relative to the cluster center, $\sim20-30\arcsec$; see
figure~\ref{fig:raddist}.  A mass $m\approx\Sigma_{\rm crit}\cdot\pi
r^2\approx 10^{14}h^{-1}M_\odot$ is interior to these arcs, much
larger than the typical masses even of cD galaxies, $\lesssim 10^{13}
M_\odot$.  Accordingly, we do not expect the cross sections for
formation of these wide separation arcs to be grossly affected by the
presence or absence of galaxies in the simulations.  On the other
hand, galaxies can strongly affect the cross section for formation of
smaller separation arcs.  To illustrate, we plot in
figure~\ref{fig:withgal} the distribution of arc radii produced by
clusters from the $z=0.3$ output, lensing sources at $z_s=2$.  
We artificially added mass concentrations to the cluster centers 
in two ways: as point masses of varying mass, and as galaxies with
fixed velocity dispersion (350 km/s) and varying mass.  It is
apparent from the figure that mass is a less important determinant of
arc properties than central mass concentration, here parameterized by
velocity dispersion.  Galaxies of different masses (e.g. $3\times
10^{12}$ and $10^{13} h^{-1}M_\odot$) but the same velocity dispersion
produce nearly indistinguishable effects on the radial distribution.
Increasing the central mass concentration can greatly increase the
number of giant arcs forming at small radii, illustrated by the curves
for which point masses were added to the central pixel.  However, it
would appear that arc statistics at very large radii ($\gtrsim
20\arcsec$) are unaffected by these central mass concentrations.  Even
at moderate radii ($\gtrsim10\arcsec$), the difference between the
statistics for pure dark matter simulations and those including
reasonably sized galaxies is not dramatic.

Accordingly, we expect that our pure dark matter simulations should
provide a reasonably accurate estimate for the optical depth for wide
separation giant arcs in the $\Lambda$CDM model.  However, absent a
prescription for galaxy formation, we cannot make reliable
predictions for the numbers of intermediate separation
($r\lesssim5\arcsec$) arcs, which will prove to be a limitation in the
interpretation of high redshift giant arc surveys.

\begin{figure}
\centerline{\includegraphics[width=0.4\textwidth]{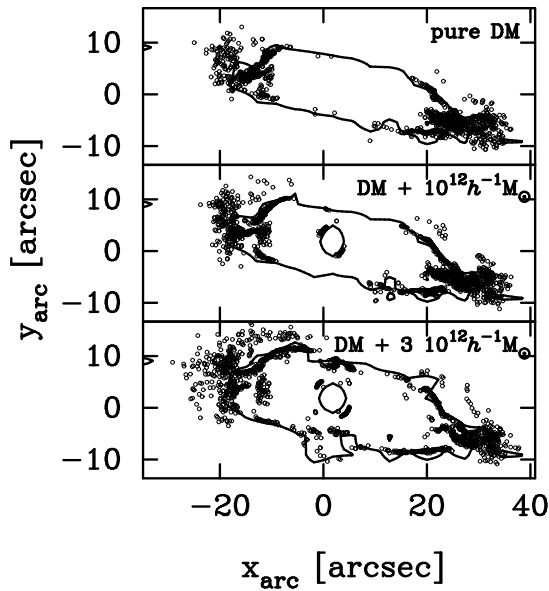}}
\caption{Angular distribution of giant arcs for one cluster
  projection.  The panels correspond, from top to bottom, to results
  from the DM simulation with added central point masses of
  0,1, and 3 $10^{12}h^{-1}M_\odot$ respectively.  The solid lines are the
  critical curves, and symbols show the centers of giant arcs produced
  in the Monte Carlo.
  \label{fig:angexample}
}
\end{figure}

Although cD galaxies do not dramatically change the wide separation
arc cross section, they can have important effects on giant arc
properties.  For example, \citet{williams99} have argued that massive
($M\sim3\times 10^{12}M_\odot$) galaxies are required at the centers
of most low redshift arc-bearing clusters in order to explain observed
arc widths.  Central galaxies also affect the angular distribution of
long, thin arcs about the cluster center. 
As an example, we plot in figure~\ref{fig:angexample} the
angular distribution of giant arcs about one particular cluster, as a
function of the mass of the central cD galaxy painted onto the surface
density.  In the top panel, corresponding to pure DM with no added
galaxy, the giant arcs tend to form only at the ends of the critical
line.  As the mass becomes more centrally concentrated, the critical
lines do not change dramatically in size, however giant arcs begin to
form all around the critical lines, instead of only along the cluster
major axis.

\begin{figure}
\plotone{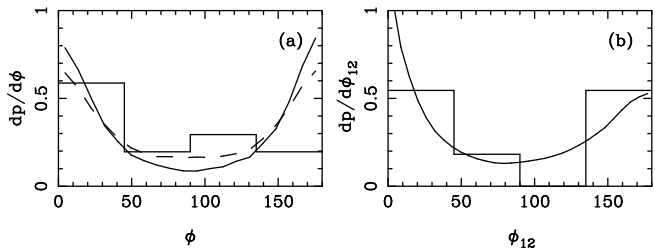}
\caption{Distribution of arc position angles from $z=0.3$ GIF clusters.
  The left panel shows arc position angles relative to the
  major axis of the projected dark matter distribution, for pure DM (solid)
  and DM + $3\times 10^{12} h^{-1}M_\odot$ (dashed).  The histogram shows
  the same distribution for the EMSS arcs. The right panel shows the 
  distribution of relative position angles between pairs of arcs in
  multi-arc systems, $\phi_{12}=\phi_1-\phi_2$, for DM only. The histogram
  shows the $\phi_{12}$ distribution for the three EMSS clusters showing
  multiple giant arcs.
  \label{fig:dndphi}
}
\end{figure}

This is shown more quantitatively in figure~\ref{fig:dndphi}, which
plots the angular distribution of arcs about the cluster center,
relative to the cluster major axis.  About 80\% of giant arcs form
at position angles within $45^\circ$ of the cluster major axis in the
pure DM simulations.  The large anisotropy in the
angular distribution is diminished somewhat by the addition of the
central galaxy; nevertheless we expect a residual anisotropy to be
present for realistic galaxy masses.  For example, adding a central
mass of $3\times 10^{12} h^{-1}M_\odot$ decreases the fraction of arcs
within $45^\circ$ of the major axis to $\sim 70\%$. 
Under the assumption that the
cluster major axis can be determined from the brightest central galaxy
position angle, this prediction can be tested using clusters even
which have only one arc.  This anisotropy signal may be washed out by
misalignments between the BCG orientation and halo orientation,
however for multi-arc clusters the anisotropy should also be manifest
in the distribution of relative position angles of arcs,
$\phi_{12}=\phi_1-\phi_2$, also shown in figure~\ref{fig:dndphi}.

This behavior can be easily understood, as outlined in more detail in
the appendix. The caustic structures of the pure dark matter case often
take the form of ``lips'' catastrophes, while the addition of the
central mass transforms these into the more familiar radial and
tangential caustics. 
Lips caustics tend to form giant arcs only at the ends of the critical
line, while tangential caustics can form giant arcs at a wide variety of
position angles. 

Since we have no reliable model for populating CDM halos with
galaxies, we cannot at present make predictions regarding the angular
distribution of arcs around clusters.  However, our crude treatment of
central galaxies seems to indicate that typical cD
galaxies do not fully isotropize the arc angular
distribution.  It will therefore be interesting to measure the angular
distribution with upcoming wide-angle, deep lensing surveys which
should find numerous multi-arc clusters. If clusters have lips, then 
multiple arcs in a cluster should tend to form along the cluster major
axis, at the ends of the critical curve. 

\subsection{Numerical resolution} \label{sec:res}

We argued above that our dissipationless N-body simulations cannot be
used to make reliable predictions for arc statistics at small radii,
while at large radii $(d>10-20\arcsec)$, neglect of baryonic physics
should not change our results significantly.  Even at large radii,
however, another possible source of error is particle noise in the
simulation. Collapsed objects represented by low
numbers of particles will suffer artificially from two-body relaxation
effects, which can significantly affect the inner structures of
simulated halos. \citet{power03} find that halos satisfying 
$t_{\rm relax}(r)=t_{\rm dyn} N(<r)/[8\log N] > 0.6 H_0^{\,-1}$
may safely be assumed to be free of relaxation effects.  We are
interested in measuring arc properties at radii $r\sim50-100
h^{-1}$kpc for objects of mass $M\gtrsim 10^{14.5} h^{-1} M_\odot$.  
For such objects, the \citeauthor{power03} bound implies a minimum of
$N(r_{\rm Ein})\gtrsim 1500$ particles representing the profile
interior to the arc radius, or a particle mass not exceeding
$m_p\lesssim 4\times 10^9 h^{-1} M_\odot$.  For comparison, the
$\Lambda$CDM GIF simulations have particle masses of $1.4\times
10^{10} h^{-1} M_\odot$, in excess of this bound.  Hence, it is
unclear from this simple analysis if our results will be affected by
particle noise.  Two-body relaxation can also become important if the
force softening is insufficient.  \citeauthor{power03} argue that a
force softening length $\epsilon\approx 4 r_{\rm vir}/\sqrt{N_{\rm vir}}$ is
optimal.  For our example halo this works out to about 
$\epsilon_{\rm opt}\approx 16 h^{-1}$kpc; while in comparison the GIF 
$\Lambda$CDM simulations use a softening of $20 h^{-1}$ kpc.

\begin{figure}
\plotone{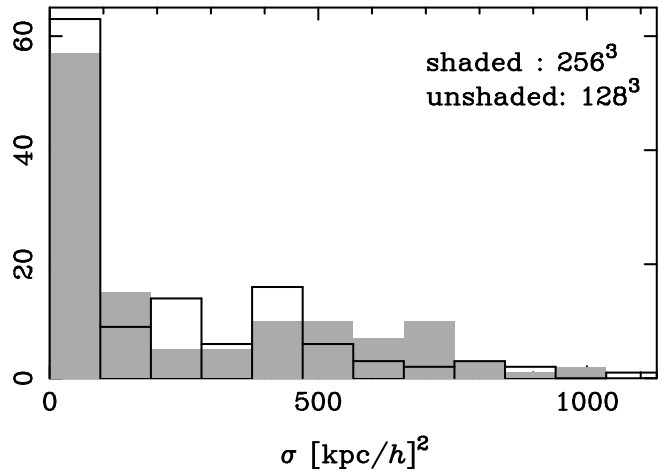}
\caption{Histogram of cross-sections for the same $M=4\times10^{14}
  h^{-1} M_\odot$ cluster at $z=0.3$ simulated at two different
  resolutions.  The shaded histogram corresponds to the high
  resolution simulation, with a mean cross section for producing
  $l/w>10$ arcs at $d>10\arcsec$ of
  $\langle\sigma\rangle=210 (h^{-1}{\rm kpc})^2$.  The unshaded
  histogram corresponds to the low resolution simulation, which gives 
  $\langle\sigma\rangle=250 (h^{-1}{\rm kpc})^2$.
  \label{fig:res}
}
\end{figure}

To test whether such resolution effects will significantly impact on
our conclusions, we perform the following convergence test.  We
perform two simulations of a $(100 h^{-1}{\rm Mpc})^3$ volume at two
different resolutions, using the TreePM code\footnote{see 
  http://www.astro.princeton.edu/$\sim$bode/TPM/index.html} 
(kindly made publicly available by P.\ Bode and J.\ P.\ Ostriker).
The low resolution simulation uses $128^3$
particles with mass $m_p=3.6\times 10^{10} h^{-1} M_\odot$ and force
softening length $\epsilon = 20 h^{-1}$ kpc, while the high resolution
simulation has $256^3$ particles of mass $m_p=4.5\times 10^9 h^{-1}
M_\odot$ and softening $\epsilon = 10 h^{-1}$ kpc.  To ensure that the
same objects would form in both simulations, they were initialized
using the same random phases for all modes common to the two boxes, while
the higher resolution run contained additional high-$k$ power.  We
then compared individual clusters from each simulation.  One example
is shown in Figure~\ref{fig:res}; the other clusters we checked showed
similar levels of agreement between the two resolutions.  We therefore
conclude that the resolution of the GIF simulations is adequate for
our purposes.

\section{How Many Luminous Giant Arcs?}

Following the considerations of \S\ref{galaxy}, we will count in our
statistics only those arcs with length-to-width ratios $r\geq10$, and
distances relative to the cluster center of $d\geq 10\arcsec$ in order
to minimize the uncertainties from the galaxy contribution.  

The optical depth for forming an arc can be derived for each simulation
volume as a sum over all clusters in the box:

\begin{equation}
\frac{d\tau}{dz_l}(z_s) = \frac{dV_{\rm co}/dz_l}{L_{\rm box}^3}
\sum_{j=1}^{N_{clus}} {\bar{\sigma}_j(z_s) \over d_A^2(z_s)}
\end{equation}

The total optical depth as a function of source redshift $\tau(z_s)$ is then 
simply a sum over simulation volumes. The number of arcs
expected per square degree is then given by
\begin{equation}
N_{arcs} = \int_0^\infty dz_s \tau(z_s) {dN_{source} \over dz_s} \quad ,
\end{equation}
where $dN_{source}/dz_s$ is the source density that could form observable
arcs. 

\begin{figure*}
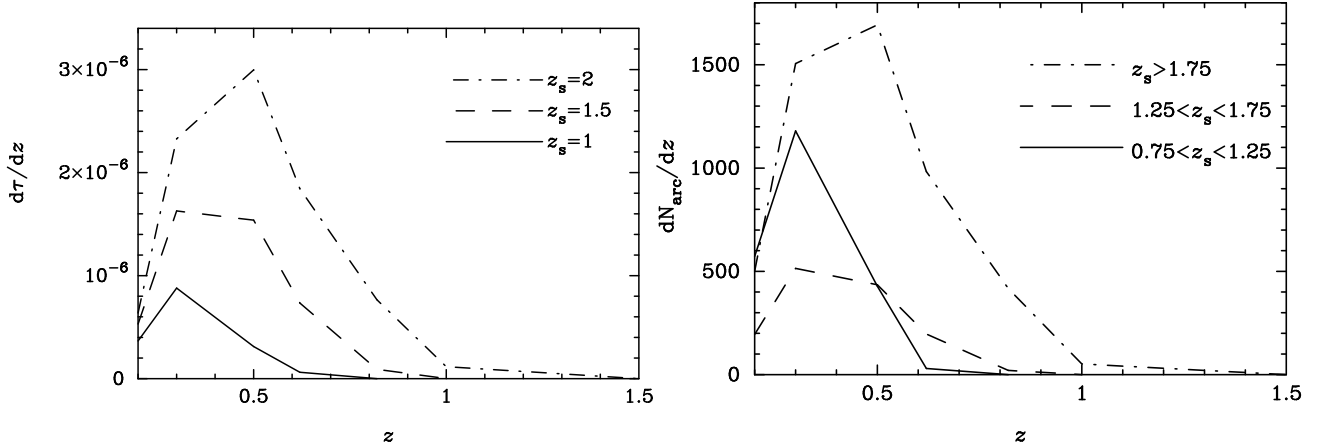

\centerline{\plotone{tauz.eps}\plotone{numz.eps}}
\caption{The left panel shows the optical depth for arcs with
  length/width $\geq10$ and distance from cluster center
  $d>10\arcsec$, as a function of redshift for GIF simulations. The
  right panel shows the number of arcs with total integrated magnitude
  $V<22$ as a function of lens redshift, for three different source
  redshift bins.  The declining number density of sources with
  increasing redshift somewhat counteracts the increase of optical
  depth with redshift.
  \label{fig:tauz}
}
\end{figure*}

In figure \ref{fig:tauz} we plot the contribution to the giant arc
optical depth from different lens redshifts, for three different
source redshifts.  As \citet{wambsganss03} have stressed, the optical
depth increases with source galaxy redshift: we find optical depths of
0.25, 0.70, \& $1.4\times 10^{-6}$ for source redshifts $z_s=1$, 1.5,
and 2 respectively.  However, we find that the variation with redshift,
while substantial, is not as strong as that reported by \citet{wambsganss03}.

The source density term is of equal importance to the optical
depth term.  To estimate the number density of galaxies as a function
of redshift and limiting magnitude, we employ the photometric
catalogs of \citet{fontana00}, compiled from galaxy counts from the
Hubble Deep Fields.  We divided source galaxies into three bins:
$0.75\leq z_s < 1.25$, $1.25\leq z_s < 1.75$, and $z_s > 1.75$.  

Magnification bias may or may not be important. In a strictly
flux-limited survey, magnification bias would be important, but 
there is no magnification bias in the limit where all sources
are resolved and surface brightness is the limiting concern. For
giant arc surveys, neither regime is appropriate. Arc detection
is closer to being limited by surface brightness than by integrated
flux, but adaptive smoothing can modify this. The observed source 
distribution is expressed in terms of integrated flux, but 
ground-based surveys may be missing distant compact sources because
seeing can reduce their apparent surface brightness. In what follows
we include magnification bias when comparing to catalogs that 
are nominally magnitude-limited. 

\begin{deluxetable}{lccccc}
\tablecaption{Survey statistics\label{tab1}}
\tablehead{
\colhead{Survey} & \colhead{redshifts} & \colhead{area} &
\colhead{depth} & \colhead{$N_{\rm obs}$} & \colhead{$N_{\rm pred}$}
}
\startdata
EMSS$^1$ & $0.15<z\lesssim0.6$ & $\sim360$ & $V<22$ & 8 & 8.2 \\
LCDCS$^2$ & $0.5<z<0.7$ & 69 & $R<21.5$ & 2 & 1.2 \\
RCS$^3$ & $z<0.6$ & 90 & $\mu_R<24$ & 0 & 2 \\
 & $z>0.6$ &  &  & 4 & 1 
\enddata
\tablecomments{Survey areas are given in square degrees.  For the
  LCDCS, quoted arc statistics are for all radii, with no cut on arc
  radius, while RCS statistics are for arcs in their primary sample
  with $d>10\arcsec$.  References are: (1) \citet{luppino99}, (2)
  \citet{zaritsky03}, (3) \citet{gladders03}. 
}
\end{deluxetable}

\subsection{The EMSS Sample}
\label{emss}

In comparing to observations it is essential that the effective
area of the survey is well-known. For the EMSS, calculating  the effective
survey area is non-trivial. The survey area is a strong function of
limiting flux \citep{henry92}. For each cluster detected in the EMSS
we found the sky area $A_i$ over which that cluster could have been detected.
An estimate of the cluster surface density 
is then given by  $\sum 1/A_i$ clusters per square degree. The
EMSS lensing sample of \citet{luppino99} is defined by clusters
with $L_x>2\times 10^{44}$erg s$^{-1}$ and $z>0.15$. 
Using the \citet{henry92} coverage, we find that the surface density of EMSS
clusters is one cluster per 9.6 square 
degrees,  a total coverage for the 38 clusters of 360 square degrees. 
The earlier sub-sample of 16 clusters, used in \citet{lefevre94}, 
is defined by  
$L_x>4\times 10^{44}$erg s$^{-1}$ and $z>0.2$, for which the surface
density is one cluster per 20 square degrees.

Of the 38 most massive EMSS clusters, 8 show giant luminous arcs. 
Extrapolating over the entire sky we would therefore expect roughly 900 giant
arcs.  The sub-sample used by \citet{lefevre94} showed 6
clusters with giant arcs, which translates into one arc-inducing cluster
every 53 square degrees or roughly 800 giant arcs on the entire sky.
The consistency of these two values supports our estimate for the area
covered by EMSS.

We can roughly estimate the expected number of arcs by taking the
optical depths from figure~\ref{fig:tauz}
and multiplying by an appropriate source
galaxy density.  Assuming that arcs with length/width ratios
$r\geq10$ are magnified by about 2.5 magnitudes, and noting that
\citet{lefevre94} counted arcs with integrated magnitudes $V<22$, we
find galaxy densities for $V<24.5$ of 7.4, 1.4, and 2.6 arcmin$^{-2}$
for our three redshift bins from the Hubble Deep Fields.  Multiplying
by the corresponding optical depths (0.25, 0.70, \& $1.4\times
10^{-6}$) gives a total of $\sim940$ arcs expected over the full
sky.  An alternative estimate for the number of arcs may be obtained
not by assuming arcs are magnified by $\sim2.5$ magnitudes, but instead
convolving the calculated arc magnification distribution with the
observed luminosity function.  This estimate would be valid in the
limit that low surface brightness arcs with total integrated
magnitudes brighter than the detection threshold are observable.
For a threshold integrated magnitude of
$V<22$, the HDF galaxy counts reported by \citet{fontana00} give
$\sim1400$ giant arcs with $r\geq 10$ located $d\geq 10\arcsec$ from
the cluster center, over the full sky.

Apparently, there is no excess in the number of giant arcs observed
around EMSS clusters compared to expectations from the $\Lambda$CDM
model.  It appears that the solution to what was thought to be an
order of magnitude problem is in a conspiracy of small effects: a
larger effective area for the EMSS than estimated in B98 (factor of
2), a higher source density (factor of 2), and a somewhat higher
redshift-averaged optical depth (factor of 2).

\subsection{LCDCS Sample}

\citet{zaritsky03} surveyed clusters in the range $0.5\leq z\leq0.7$
over a 69 deg$^2$ field.  In this area, they found 
two giant arcs with $R<21.5$ and length $L>10\arcsec$, and a number of
additional arcs and arc candidates with which they do not include in
their statistics. 
A straightforward comparison with our expected statistics is
complicated by the fact that both of these arcs are located close
($\lesssim10\arcsec$) to cluster galaxies.  As mentioned in
\S\ref{galaxy}, the optical depth for such close-in arcs is strongly
affected by the presence of galaxies which are neglected in the pure
dark matter simulations we have used.

Nevertheless, we can try to make some rough estimates for the expected
giant arc rate.  Dropping the $d>10\arcsec$ cut we imposed for the
EMSS sample, we find that over the range $0.5<z<0.7$, the GIF clusters
produce 0.7 arcs in the 69 deg$^2$ surveyed in the LCDCS.  If we paint
galaxies with $\sigma=350$ km/s and $M=3\times 10^{12}h^{-1}M_\odot$
onto the clusters, the expected number increases to 1.2 giant arcs.

Accordingly, it appears that the giant arc incidence measured by the
LCDCS is consistent with predictions from the $\Lambda$CDM model.
However, \citeauthor{zaritsky03} emphasize that their measured giant
arc statistics should be viewed as a lower bound on the true
incidence; a less conservative cut on arcs in their sample could lead
to a statistically significant discrepancy.  On the other hand,
\citeauthor{zaritsky03} also note that their estimated cluster number
density is $4.5\times$ higher than that of the EMSS at these
redshifts, which could artificially enhance the LCDCS arc rate
relative to the full sky average.

\subsection{RCS Sample}

\citet{gladders03} survey 90 deg$^2$ for giant arcs, counting only
arcs with surface brightness in $R_C$ of $\mu_R < 24 \mbox{ mag 
arcsec}^{-2}$.  In their
primary sample, \citeauthor{gladders03} detect a total of 5 arc
candidates, four of which have length-to-width ratios $r\geq 10$.
Oddly, all five of the arc-candidate clusters in their primary sample
have high redshifts, $z\geq 0.64$.  From the EMSS statistics, we would
expect on average $\sim2$ arc-producing clusters in the range
$0.15\leq z\leq0.6$ over a 90 deg$^2$ field, so the lack of such
clusters in the RCS sample is unusual but not significantly
discrepant.  

Since \citeauthor{gladders03} impose a surface brightness cut on their
arc candidates, to compare with their statistics we need to know the
source galaxy number density as a function of surface brightness.
\citet{fontana00} do not give surface brightnesses for the galaxies in
their Hubble Deep Field catalog, but they do provide magnitudes and
half-light radii for galaxies observed in the NTT Deep Field.  From
their catalog, we derive number densities of galaxies with $\mu_R <
24 \mbox{\, mag arcsec}^{-2}$ of 6.9, 1.7, and 9.0 arcmin$^{-2}$ for our
three source redshift bins.  \citeauthor{fontana00} do not give
estimates for their incompleteness at these surface brightnesses.
From figure~\ref{fig:tauz}, we can see that GIF simulation lenses at
$z>0.6$ primarily produce arcs from sources at $z_s\gtrsim 2$, so for
simplicity we will only count these sources in our statistics.  

From the GIF simulation outputs at $z=0.62$ and $z=0.82$, we would
expect $\sim 1$ arc in the RCS field with length-to-width ratios
exceeding 10, and distances from the cluster center exceeding
$10\arcsec$. \citeauthor{gladders03} find four such arcs around three
clusters in their primary sample.  In addition, they also report
another giant arc at smaller radii. Dropping the cut on $d\geq
10\arcsec$ increases the expected number of arcs with length-to-width
ratio $r\geq 10$ to $\sim2$, while painting galaxies with $\sigma=350$
km/s onto the clusters increases the expected number of arcs to
$\sim5$.

In addition to the arc candidates included in their primary sample,
\citet{gladders03} also report the discovery of three arc-producing
clusters found in a follow-up survey of clusters with redshifts
$z>0.95$.  Two of the clusters have arcs at small radii, however the
third, RCS 2319.9+0038, has multiple arcs at $\gtrsim 10\arcsec$.

The RCS statistics may indicate an excess of giant
arcs observed at high redshifts, in contrast to the low-redshift
agreement found between EMSS and the expected GIF arc statistics.
The excess appears to be due to the presence of extremely massive
and/or concentrated
clusters in the RCS fields which are not present in the GIF volume.
For example, RCS 2319.9+0038 has $\approx 10^{14.2} h^{-1} M_\odot$
enclosed within the central $20\arcsec$, while the most massive
cluster in the $z=1$ GIF volume has less than twice that mass inside
its virial radius!  The high fraction of multiple-arc systems in the
RCS sample further supports the notion argued by
\citeauthor{gladders03}\ that a few ``super-lenses''
are responsible for much of the optical depth, although a
full treatment of multiple arc systems must take into consideration
source clustering, which is certainly important in at least some
multi-arc clusters \citep[e.g][]{sand02}.  

\begin{figure}
\plotone{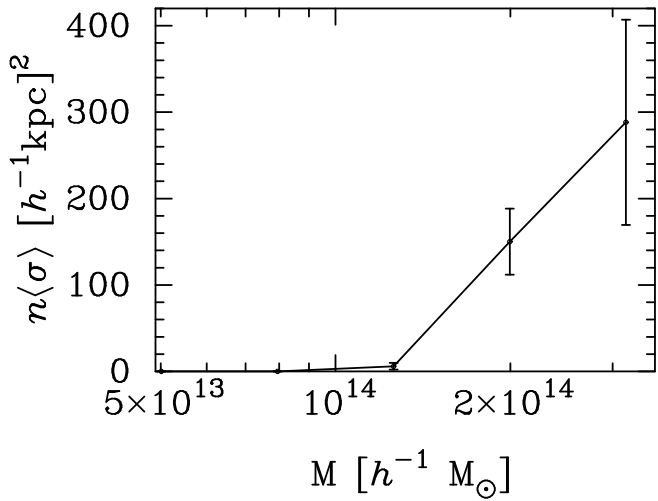}
\caption{Number weighted cross-section at redshift $z=0.82$ for
  sources at $z_s=2$. The strong rise out to the largest masses may
  indicate that the dominant lenses are not represented in the GIF
  volume at high redshifts.
  \label{sigm0.82}
}
\end{figure}

Unfortunately, these rare clusters are
precisely the clusters for which the small volume $(\sim10^6 h^{-3}
\mbox{Mpc}^3)$ we use is deficient in describing the much larger volume
($>10^8 h^{-3}\mbox{Mpc}^3$ comoving) probed by the RCS sample.
Massive objects like RCS 2319.9+0038, which evidently occur only once
or so per RCS volume, will almost certainly never appear in a volume two
orders of magnitude smaller.  Indeed, examination of the giant arc
optical depth as a function of mass at the relevant redshifts
(Figure~\ref{sigm0.82}) indicates that we are missing the objects
which dominate the optical depth. Accordingly, we cannot say at present
whether the RCS results are consistent or not with the
$\Lambda$CDM model.

\section{Comparison with Previous Work}

We find rough agreement with the optical depths reported by B98
for giant arcs with length-to-width ratios $r\geq10$.  Nevertheless,
we do not find that this optical depth leads to an order of magnitude
deficit in the number of arcs expected in EMSS clusters compared to
the observed number.  Part of the difference between our conclusions
and those of B98 stems from the difference in estimates for cluster
number counts.  We have estimated this using directly the EMSS sky
coverage reported in \citet{henry92}, while B98 assume a number
density for EMSS clusters corresponding to about twice the sky density
we have estimated.  In addition, B98's estimate for the source density
is lower than our estimate (based on counts from the Hubble Deep
Fields) also by a factor of two.  
This combined with our somewhat larger optical depths 
accounts for the difference in our conclusions.  One puzzling
difference between our results is that B98 find a much steeper dependence
of optical depth on arc length-to-width ratio.  We have quoted
statistics for arcs with $r\geq10$.  For lower thresholds, e.g. $r\geq
7.5$, our optical depths increase by a factor $\sim2$.  In contrast,
B98 find that the optical depth for $r\geq 7.5$ is nearly an order of
magnitude larger than that for length-to-width ratio of 10.
Theoretically, one might expect the cross section to scale like
$\sigma\propto l_c \delta(\mu)$, where $l_c$ is the caustic length and
$\delta$ is the distance from the caustic for which sources are
magnified by $\mu$.  Near folds, the magnification scales like
$\mu\propto\delta^{-1/2}$, while near cusps the magnification scales
like $\mu\propto\delta^{-1}$ \citep{sef}.  Naively, assuming
$r$ is proportional to $\mu$ near critical lines, one might
expect the cross section to scale like $\sigma\propto 1/r^{1-2}$,
consistent with the moderate rise in $\tau$ with declining $r$ which
we find.  We also note that other workers \citep{oguri03,wambsganss03}
find $r$ dependence similar to ours in their simulations, and
certainly the observed counts of giant arcs do not show such a steep
dependence on length-to-width ratio.

We do not reproduce the enhanced optical depths of
\citet{wambsganss03}, and in particular we do not find as steep a
redshift dependence.  For example, their reported optical depth at
$z_s=1.5$ of $\tau=2.3\times 10^{-6}$ is a factor of 3 larger than our
value.  It is possible that part of this discrepancy could be
attributed to their use of arc magnification as a proxy for
length-to-width ratio; as shown in figure~\ref{wbo} this could
overestimate the cross section by a factor of 2-3.

The weaker dependence on source redshift is more difficult to explain.  
The main difference between our strategies was that we only included material 
within $\sim100$ Mpc
of the cluster and did not integrate along the entire light
cone. However, given the expected fluctuations of only $\sim 1\%$ of
the critical surface density, it seems unlikely that large scale 
structure could affect the
lensing cross-section significantly.  It is possible to significantly
enhance the cross-section with small fluctuations in surface density
for lenses with Einstein radii much smaller than the scale radius,
where the surface density profile is extremely shallow, but this is
not the regime in which we expect to see giant cluster arcs.  We have
explicitly checked that the addition of extra surface density at the
level of $\delta\kappa\approx 3\%$ changes the giant arc cross section
by $\lesssim 5\%$ for the massive clusters which dominate the optical
depth. 

The simulations of \citeauthor{wambsganss03} had significantly higher
force resolution than the GIF simulations:
\citeauthor{wambsganss03} used N-body simulations with
$\sim 3$ kpc force resolution, while the GIF simulations had
force resolution of 20 $h^{-1}$ kpc.  However, 20 $h^{-1}$ kpc is still
well within the typical radii at which these giant arcs form, and we
have argued in \S\ref{sec:res} that this resolution is sufficient to
calculate the total optical
depth to forming giant arcs.  At present, we do not have a
good understanding for the different behavior with redshift but our
investigations of the effects of central galaxies showed that it is
possible that the redshift behavior is sensitive to which arcs are
considered to be giant cluster arcs.

The enhanced cross-sections found by \citet{oguri03} are also puzzling,
since the main effect seems to come from considering steeper profiles in
the inner region ($\sim 10-20$ kpc), well inside typical Einstein radii 
($\sim 50-100$ kpc). Again, this could plausibly be a sensitivity in the
total cross-section to the details of the central region that is in
reality not relevant to the problem of giant cluster arcs.  

\citet{gladders03} have suggested that ``superlenses'', 
objects with low mass but large cross section, may be responsible for
much of the optical depth.  Their argument for low masses hinges on
the assumption that the cross-section for giant arcs scales linearly
with the mass, but fig.~\ref{fig:cross-section} 
shows that a more reasonable estimate is $\sigma \propto M^{3-4}$. 
With this steeper scaling it is plausible that ``superlenses'' are
primarily simply more massive clusters.  Weak lensing mass maps at
larger radii or Sunyaev-Zeldovich measurements can answer the question
of whether high-redshift lenses like RCS 2319.9+0038 are supermassive
(e.g.\ $M\sim10^{15}h^{-1}M_\odot$) or not.  If these objects have
lower masses (e.g.\ few$\times10^{14}h^{-1}M_\odot$), then their large
Einstein radii are puzzling.  \citet{meneghetti03m} and \citet{torri03}
have suggested that cluster mergers may account for a significant
fraction of the giant arc optical depth.  However, note that
the critical lines
traced out by the multiple-arc systems of \citeauthor{gladders03}\
appear relatively round, in contrast to the highly elongated critical
lines expected of low mass clusters whose giant arc cross sections
have been enhanced by ongoing mergers \citep{meneghetti03m,torri03}.
Another possibility is that ``super-lenses'' may be ordinary mass, but
super-concentrated.  An example of this may be CL0024+1654,
for which a recent combined strong- and weak-lensing analysis
\citep{kneib03} indicates a concentration $c_{200}\approx22$.  Such
high concentrations are extremely anomalous for simulated CDM galaxy
clusters \citep[e.g][]{bullock01}.  If the RCS super-lenses turn out
to have similarly deviant concentrations, and such objects are
commonplace, their existence could pose a serious challenge to our
understanding of structure formation in the $\Lambda$CDM model.

\section{Discussion and Conclusions}

At present, there appears to be no discrepancy between observed giant
arc statistics and predictions from the $\Lambda$CDM model at low
redshifts $z\lesssim 0.6$.  Imaging of EMSS clusters by
\citet{lefevre94} and \citet{luppino99} indicates that $\sim900$ giant
arcs are expected over the full sky from X-ray bright clusters at
these redshifts, while ray-tracing through clusters taken from the GIF
simulations leads us to expect $\sim1000$ giant arcs.  At higher
redshifts $z\gtrsim 0.6$, the observed abundance of close-in arcs with
$d\lesssim 10\arcsec$ observed by the LCDCS and RCS appears consistent
with expectations, when the 
effects of the central galaxies are included in the gravitational
potential.  However, there may be an excess of wide separation 
$(d\gtrsim 10\arcsec)$ giant
arcs at high redshift.  This putative excess appears to be due to
extremely massive or unusually concentrated clusters not present in
the (admittedly limited) simulation volumes we have employed.  

Looking forward, upcoming X-ray cluster surveys like MACS
\citep{ebeling01} and wide-area surveys like the CFHT Legacy
Survey\footnote{http://www.cfht.hawaii.edu/Science/CFHLS} and the 
Sloan Digital Sky Survey\footnote{http://www.sdss.org} can be
expected to improve the statistics of giant arcs on the sky.  For
example, the RCS-2 survey covers an area of 830 deg$^2$ and is
expected to produce $\sim 50-100$ new arcs (M.\ Gladders, private
communication).  Deep surveys can also begin to test CDM predictions
for giant arc properties.  As we have discussed earlier, CDM clusters
tend to produce arcs along (but aligned orthogonal to) their major
axes.  This is modified somewhat by the presence of central cD
galaxies in the clusters, however for reasonable velocity dispersions
we expect the arc angular distribution to be anisotropic.  Multi-arc
systems may provide a probe of halo triaxiality at small radii,
complementary to weak lensing studies at large radii
\citep[e.g.][]{hoekstra03}.

\acknowledgments{We thank Paul Bode and Jerry Ostriker for help with
N-body simulations and for making publicly available their TreePM
code.  We also thank Matthias Bartelmann, Mike Gladders and Massimo
Meneghetti for helpful discussions.  N.\ D.\ acknowledges the support of
NASA through Hubble Fellowship grant \#HST-HF-01148.01-A awarded by
the Space Telescope Science Institute, which is operated by the
Association of Universities for Research in Astronomy, Inc., for NASA,
under contract NAS 5-26555.  G.H.\ is supported by the W.\ M.\ Keck
Foundation. J.\ F.\ H.\ is supported by a Proctor Fellowship granted
by Princeton University.  }

\appendix
\section{On Lips and Naked Cusps}

In section \S\ref{galaxy} the addition of a centrally concentrated mass
was shown to have a significant effect on where giant arcs can form.
We can understand this behavior as follows.  On the critical curves,
the magnification matrix becomes singular, causing sources to become
highly magnified along one direction.  The highly elongated direction,
however, is not in general parallel to the critical line itself.  In
some instances, the magnified direction can be perpendicular to the
critical line, i.e. it can point along $\nabla\mu$.  Such cases clearly
are not conducive to the formation of giant arcs, as the magnification
diminishes so quickly along their lengths that these arcs are unable to
acquire `giant' dimensions.  Instead, giant
arcs form preferentially at locations where the magnified direction is
aligned with the tangent to the critical curve.

The direction along which arcs are elongated is determined by the
orientation of the local shear.  For our purposes, the shear
has two principal sources: the tidal gravity associated with the
radial density gradient, and the triaxiality and substructure of the dark
matter distribution.  The former produces shear which is azimuthally
oriented with respect to the central mass concentration, while the
latter lead to shear aligned with the principal axes of the density
distribution.  The competition between these two effects determines
whether arcs are primarily tangential or `Cartesian', which determines
whether the caustic structure is that of lips, or radial and
tangential caustics.  

N-body simulations in the cold dark matter
model tend to produce clusters which have shallow radial profiles and
strong triaxiality.  Average radial profiles exhibit $r^{-1}$ density
cusps \citep[e.g][]{nfw} which, in projection, give surface densities
rising only logarithmically at small radii.  Typical clusters
are also quite triaxial, with axis ratios on average near 0.5 with
large scatter \citep[e.g][]{thomas98,jingsuto}.  Because of this, a large
fraction of GIF clusters have lips caustics, although this depends in
part on projection.  Clusters viewed along their major axes have
larger central surface densities and less ellipticity than other
projections, meaning that clusters can have radial and tangential
caustics viewed along certain projections, and lips when viewed in
other orientations.

\begin{figure}
\plotone{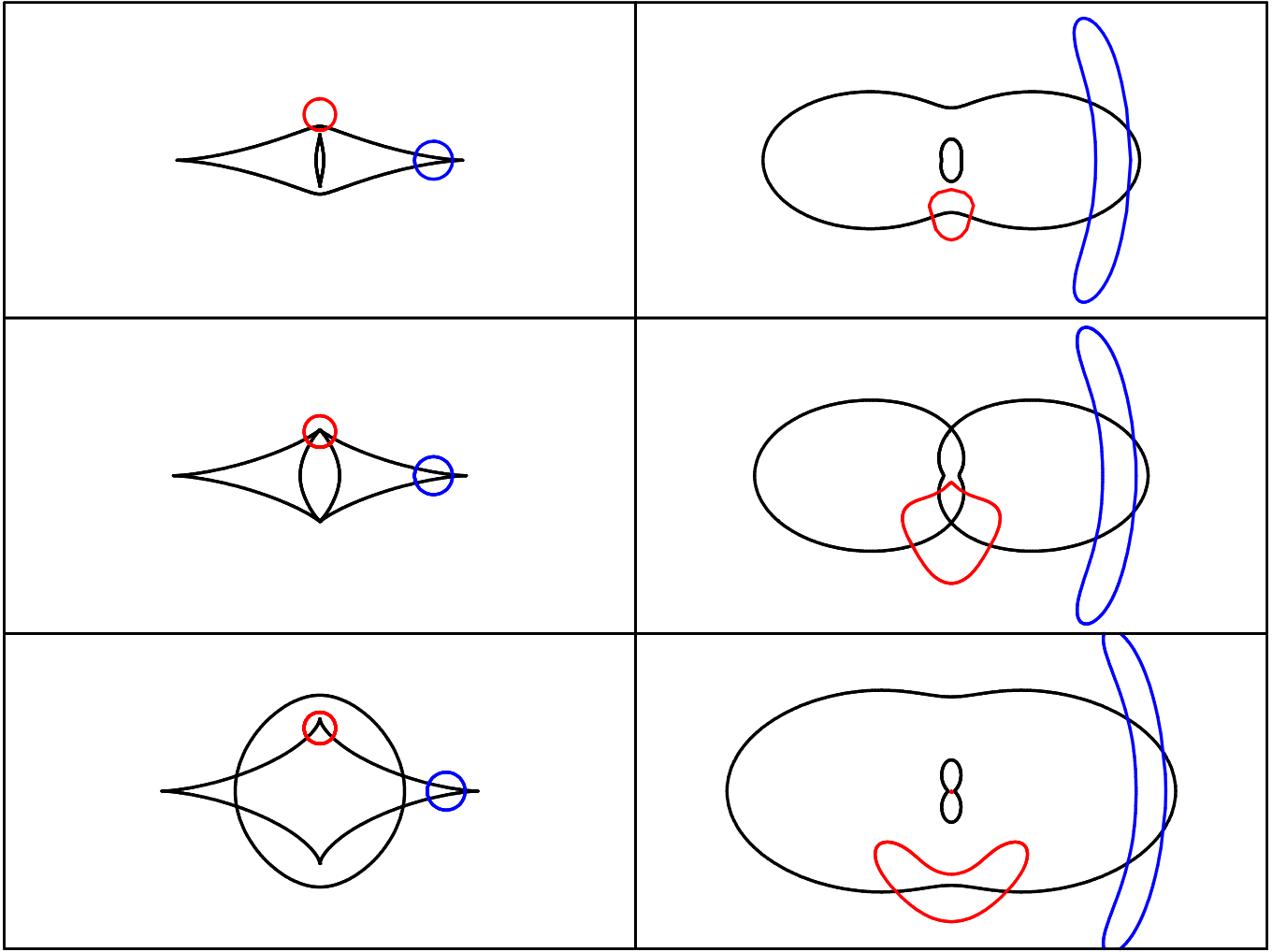}
\caption{Critical lines and caustic structure as a function of
  central mass concentration.  The top panel depicts lensing by an
  elliptical NFW profile, while the bottom two panels show the effect
  of adding increasingly massive central galaxies, modeled as
  Jaffe profiles.  Critical lines, caustics, and arcs
  were computed in this figure using the software of \citet{keeton01}.
  \label{fig:lips}
}
\end{figure}

In figure~\ref{fig:lips} we plot the caustics and critical lines for
an elliptical NFW profile, and
lensed images of representative sources.  The top panel illustrates
typical caustic patterns for the pure DM simulations.  There are two
critical lines, pointing along and orthogonal to the cluster major
axis.  However, as the lensed arcs indicate, the magnified direction
along the outer critical line tends to point up and down.  Because of
this, the corresponding caustics in the source plane take the form of
lips catastrophes, with two cusps each, corresponding to the two
locations on each critical line where the highly magnified direction
becomes tangent to the curve \citep{sef}.  This is why long,
thin arcs preferentially form at the ends of the critical lines in
this case.

The caustic structure changes as the central mass concentration
increases, shown in the bottom two panels of the figure.  
As the central mass is increased, the tangential shear grows
commensurate to the ellipticity-induced shear, and the two critical lines
approach each other.  When they meet, their corresponding caustics
meet at a hyperbolic umbilic catastrophe \citep{sef}, shown in
the middle panel of figure~\ref{fig:lips}.  Near
hyperbolic umbilics, the source becomes highly magnified in {\em both}
directions, which again is not favorable for producing long, thin
arcs.  For higher central concentrations, the critical curves become
tangential and radial; that is, their magnified directions tend to
point in the azimuthal and radial directions, for the outer and inner
curves, respectively.  The near alignment of the magnified direction
with the curve tangent, for the tangential critical line, allows long
thin arcs to form all along its length, as depicted in the bottom
panel of figure~\ref{fig:lips}.

\end{document}